\documentclass[10pt]{article}

\usepackage{amsmath}
\usepackage{amssymb}
\usepackage{soul}
\usepackage{graphicx}
\usepackage[tight,footnotesize]{subfigure}
\usepackage{cite}

\usepackage{color} 

\usepackage[labelfont=bf,labelsep=period,justification=raggedright]{caption}

\makeatletter
\renewcommand{\@biblabel}[1]{\quad#1.}
\makeatother

\date{}

\begin{document}

\begin{flushleft}
{\Large
\textbf{Does Quantum Interference exist in Twitter?}
}
\\
Xin Shuai$^{1\ast}$, 
Ying Ding$^{2}$, 
Jerome Busemeyer$^{3}$,
Yuyin Sun$^{2}$,
Shanshan Chen$^{2}$,
Jie Tang$^{4}$
\\
\bf{1} School of Informatics and Computing, Indiana University, Bloomington, IN, USA
\\
\bf{2} School of Information and Library Science, Indiana University, Bloomington, IN, USA
\\
\bf{3} Department of Psychological and Brain Sciences, Indiana University, Bloomington, IN, USA
\\
\bf{4} Department of Computer Science and Technology, Tsinghua University, Beijing, China
\\
$\ast$ E-mail: Corresponding xshuai@indiana.edu
\end{flushleft}

\section*{Abstract}
It becomes more difficult to explain the social information transfer phenomena using the classic models based merely on Shannon Information Theory (SIT) and Classic Probability Theory (CPT), because the transfer process in the social world is rich of semantic and highly contextualized. This paper aims to use twitter data to explore whether the traditional models can interpret information transfer in social networks, and whether quantum-like phenomena can be spotted in social networks. Our main contributions are: (1) SIT and CPT fail to interpret the information transfer occurring in Twitter; and (2) Quantum interference exists in Twitter, and (3) a mathematical model is proposed to elucidate the spotted quantum phenomena.

\section*{Introduction}
Shannon Information Theory (SIT)~\cite{shannon} and Classic Probability Theory (CPT) quantify the information by encoding information as symbols and ignoring their semantic aspects. Such a strategy is very successful in capturing the essential structure and dynamics of information transfer, therefore becoming the backbone of the modern telecommunication and network transmission technology. As the Web2.0 increases in popularity, our current ways of communicating are changing dramatically. Facebook has become gathering spot in our daily lives by connecting like-minded folks and establishing virtual communities to solve problems and accomplish tasks. News and comments about natural disasters (e.g. JapanÕs Tsunami) and political uprisings (e.g. Middle EastÕs antigovernment protests) travel with a lightning speed in Twitter. As a result, the focus of information transfer gradually move from its technological aspect to social aspect, where the semantic and human factor becomes so important that cannot be neglected. 

It becomes more and more difficult to explain the social information transfer phenomena using the classic models based merely on CPT and SIT, because the transfer process in the social world is highly contextualized. The influence of possible channels of flowing information in the complex and yet dynamic social networks is sophisticated enough that any classic probabilistic model might hardly satisfy. Many recent studies are devoted to social information transfer in different social networks~\cite{flickr, galuba, david, eldar} and several interesting findings have been observed. However, none of them attempted to disclose the insufficiency of SIT and CPT to interpret some aspects of social information transfer or proposed a new model that that may better serve as a supplementation to the existing classic models.

Recently Purdue University received \$25 million funding from U.S. National Science Foundation to create the Science of Information Center to move beyond Shannon theory. They aim to develop principles to encompass concepts of structure, time, space, and semantics to aid better understanding of social networks and social media behaviors~\cite{savage}. The information is no longer the 1s or 0s of binary code and its meaning can be contextually interpreted based on different temporal, spatial and semantic factors. The value of information can change dramatically over time and the flow of information can heavily depend on how people use and trust these channels. The Web is becoming more complicated and dynamic, currently there is no good way to measure how information is transferred and evolved on the Web. SIT only cares about the physical essence of the information but ignore its semantic essence.

 It is time to rethink how information is diffused through social networks so that we can better understand the essential difference between social information transfer and physical information transfer. The origin of information transfer in social networks comes from social influence, which occurs when an individual's thoughts, feelings or actions are affected by other people. Information transfer characterizes the way that a node in social network can spread information to its neighbor nodes via exerting social influence to them. Consequently, those neighbor nodes can continually influence their own neighbor nodes to further spread that information. Given the complexity in social information transfer, some well-established conclusion in SIT and CPT may not hold any more. 
 
According to SIT, there is a default assumption in information transfer that \emph{all information transfer channels are independent and there are no interferences among them}. However, this default assumption might be incorrect. For example, if a node in social network is influenced by two neighbor nodes, and tries to make decisions based on the information obtained from the two neighbors, it is natural for this node to consider the two neighbors together rather than think of them independently. In other words, the information transferred via the two neighbors might have an interference effect on each other. Such interference between different information channels is very similar to the interference between different light waves that has already been explained by Quantum Theory (QT)~\cite{young}. Grounded on a mathematical basis, QT bears the potential to model and calculate the contextual and semantic information associated with one entity, which takes into account not only the independent features of this entity but also the dependent features in different contexts and the interference of these dependent features.

%

Whether quantum-like phenomena can be spotted in social networks is the question we hope to explore. To address it, we specifically studied the information transfer in Twitter and our main contributions include:
\begin{itemize}
\item discovered the phenomenon that the amount of information transfer in Twitter NOT always monoton-ously increases along with number of channels, which is incompatible with some conclusions in SIT.
\item proposed a quantum version of \emph{q-attention} model that is capable of mathematically interpreting the conflict between decreased information transfer and increased channels 
\end{itemize}



\section*{Methodology}

\subsection*{Information Transfer in Social Network}
\begin{figure}[!ht]
	\centering
\subfigure[]{
	\includegraphics[scale=0.35]{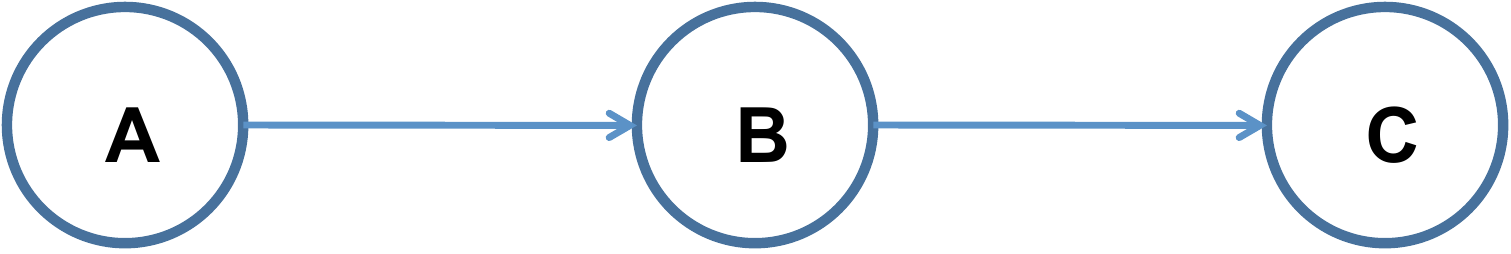}
	 \label{fig:a}
}
\subfigure[]{
	 \includegraphics[scale=0.35]{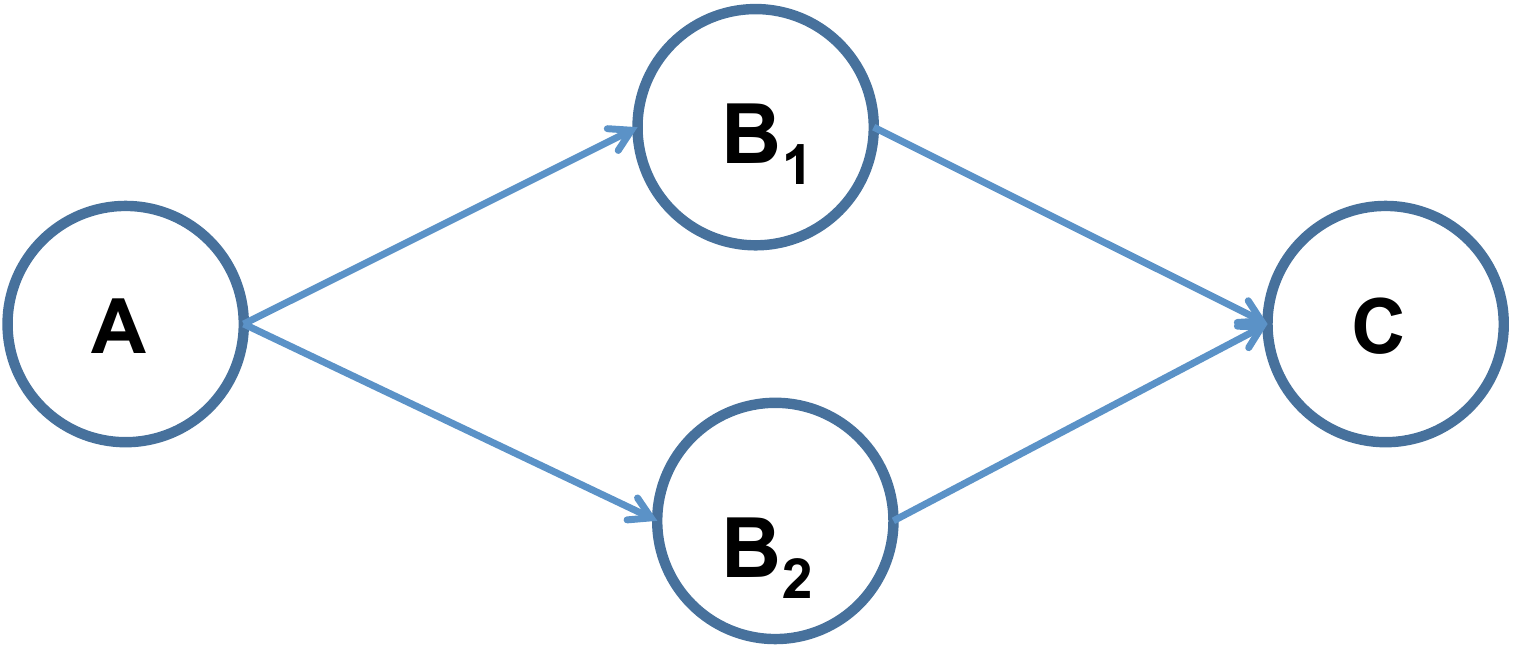}
	  \label{fig:b}
}
\subfigure[]{
	  \includegraphics[scale=0.35]{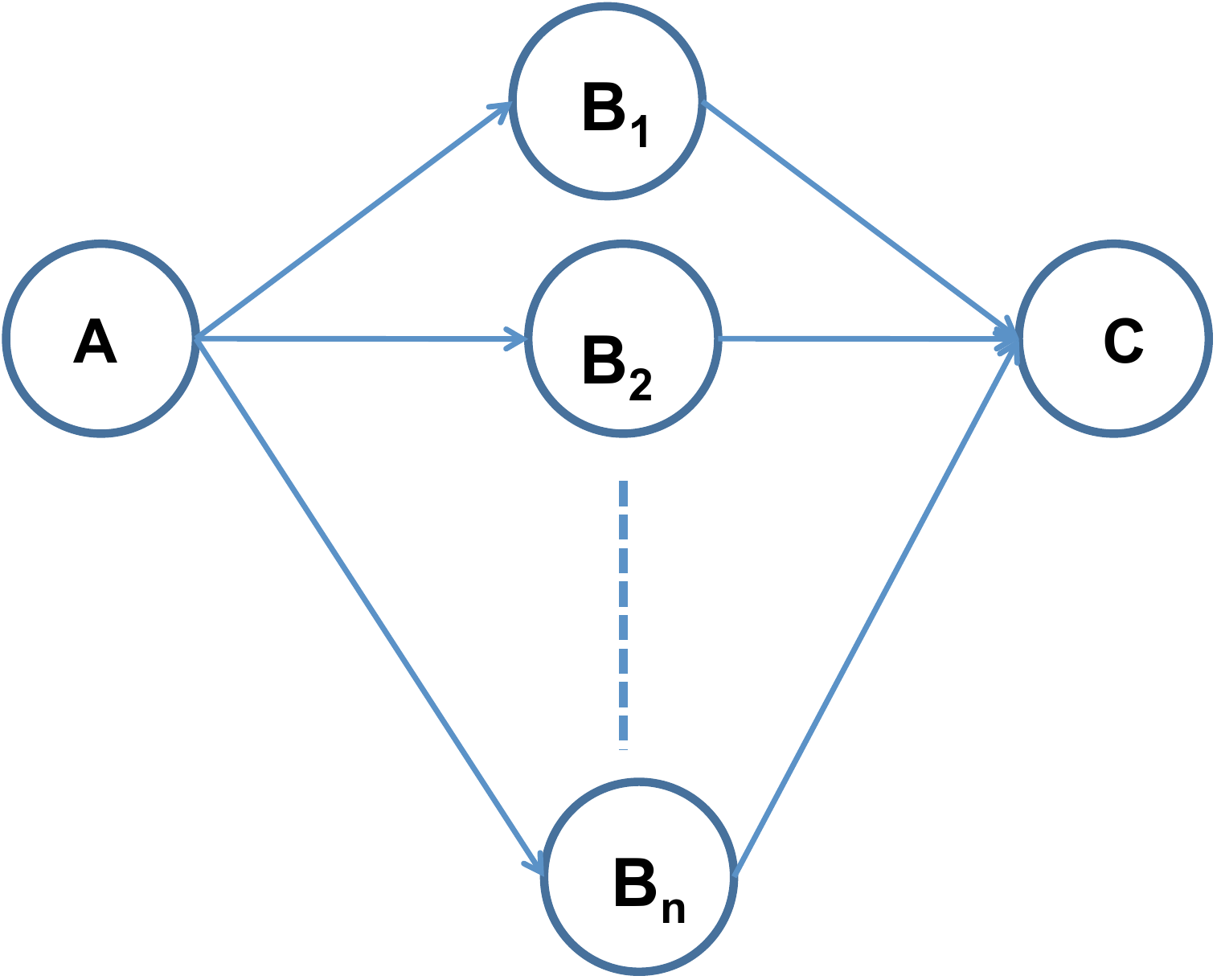}
	  \label{fig:c}
}
\caption{\bf {Information transfer patterns}}
\label{fig:pattern}
\end{figure}

In a social network, there're no clearly defined information source, channel and receiver, because each agent in the network is a node. According to different situations, each node can act as a source, or a channel or a receiver. Figure~\ref{fig:pattern} shows different information transfer patterns: 1) one-channel information transfer pattern: $A$ is the source while $C$ is the receiver, and information flows from $A$ to $C$ via a channel created by an intermediate node $B$; 2) two-channel information transfer pattern: $A$ is the source while $C$ is the receiver, and information flows from $A$ to $C$ via two channels created by intermediate nodes $B_1$ and $B_2$, respectively; and 3) n-channel information transfer pattern: $A$ is the source while $C$ is the receiver, and information flows from $A$ to $C$ via a number of channels created by a list of  intermediate nodes $B_1$, $B_2$,$ ... $, $B_n$.  A simple research question can be specified as: \emph {How does the total amount of information transferred from $A$ to $C$ change with the increase/decrease of the number of available channels?}
 

\subsection*{Information transfer in Twitter}
Twitter users send and read messages called tweets. One user can follow other users and retweet their tweets. Such retweeting relationships connect Twitter users and form the social network where information is transfered via retweeting, as identified by $RT @username$ in tweets. Assume $A$ is a Twitter user who has two followers $B_1$ and $B_2$.  $B_1$ and $B_2$ have some common followers and $C$ is one of them. One tweet coming from user $A$ can be retweeted by both $B_1$ and $B_2$. If $B_1$ or $B_2$ retweets $A$'s tweets, their tweets contain \emph{RT @A}. Here, we assume that $C$ does not directly follow $A$ and can only see $A$'s tweets retweeted by $B_1$ or $B_2$. So, $C$ can decide whether to further retweet $A$'s tweets from $B_1$ or $B_2$. Those retweets posted by $C$ contain the sign as: $RT @B_1: @A$ or $RT @B_2: @A$ which means that $C$ retweets $A$'s tweets via $B_1$ or $B_2$. 
%
%


Now we define the retweeting probability to measure the amount of information transferred from one user to another through retweeting, directly or indirectly. If $A$'s tweets are retweeted by $B$, then the retweeting probability is the percentage of $A$'s tweets retweeted by $B$. For example, if $A$ posted 100 tweets and 20 of them are directly retweeted by $B$. Then, the retweeting probability of $B$ from $A$ is 20/100 = 0.2, marked as {\bf P}$(B|A) = 0.2$. Consider the situation where an intermediate node $B$ exists between $A$ and $C$ (Figure~\ref{fig:pattern}(a)). If $A$ posted 100 tweets and 20 of them are retweeted by $B$. Of the 20 retweets, 5 are further retweeted by $C$. Then the retweeting probability of $C$ from $A$ via $B$, is 5/100 = 0.05, marked as {\bf P}$(C|A; B) = 0.05$. Note that it is different from the case that $C$ directly retweets $A$, in which no intermediate node $B$ exists.

If the number of channels between $A$ and $C$ increases to two(Figure~\ref{fig:pattern}(b)), we can define three retweeting probabilities: {\bf P}$(C|A;B_1)$ indicates the probability that $A$'s tweets are retweeted by $B_1$ and further retweeted by $C$ as if $B_2$ does not exist; {\bf P}$(C|A;B_2)$ indicates the probability that $A$'s tweets are retweeted by $B$ and further retweeted by $C$ as if $B_2$ does not exist; {\bf P}$(C|A;B_1,B_2)$ indicates the probability that $A$'s tweets are retweeted by either of $B_1$ or $B_2$, and further retweeted by $C$. 

In general, as is shown in Figure~\ref{fig:pattern}(c), we can define $n$ separate retweeting probabilities (e.g., ${\bf P}(C|A;B_1), \newline{\bf P}(C|A;B_2),...,{\bf P}(C|A;B_n)$) and an overall retweeting probability, i.e. {\bf P}$(C|A;B_1,B_2,...,B_n)$. The relationship between {\bf P}$(C|A;B_1,B_2,...,B_n)$ and $n$ separate probabilities, i.e., ${\bf P}(C|A;B_1),...,{\bf P}(C|A;B_n)$ is of our interest.

\subsection*{Q-attention Model}
\begin{figure}[!ht]
\begin{center}
\includegraphics[scale=0.25]{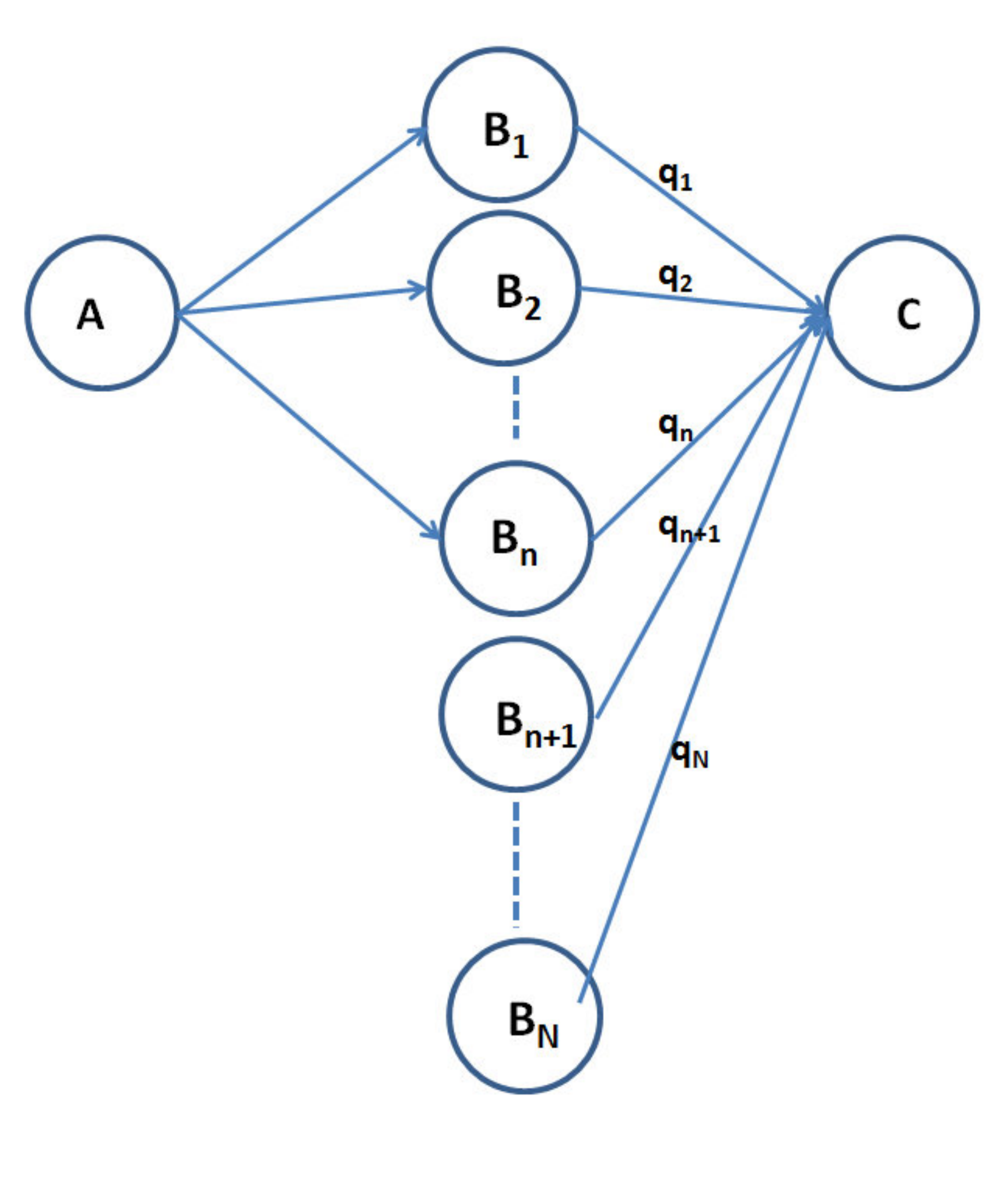}
\end{center}
\caption{
\bf {q-attention model}  
}
\label{fig:qt_model}
\end{figure}

 To determine the relationship between retweeting probability and the number of channels, we proposed a \emph{q-attention} model based on CPT by taking into account human cognition (Figure~\ref{fig:qt_model}). The \emph {q-attention} model originates from Batchelder's work~\cite{batch} presenting a family of processing models for the source-monitoring paradigm in human memory, but is specifically modified to cater to our information transfer research in Twitter. The model is designed for the situation in which agent $A$ posts a tweet that is received by a group of agents $B_{j}$, $j=1,2,...n$, who may or may not choose to retweet it to agent $C.$ There is no direct connection between $A$ and $C$ and the tweets from $A$ can only reach $C$ through one of the agents $B_{j}.$ We examine this probability as a function of the number, denoted $n,$ of agents $B_{j}$ who both see $A$'s tweets and retweet them to $C.$ It is true that there is no universal way to build a classic model based on CPT and human cognition for information transfer
 and our \emph{q-attention} model is just one of many possible models.


We assume that agent $C$ has some limits on capacity for the number of tweets that can be considered. Consequently agent $C$ can only pay attention to a total number of $N$ intermediate agents forming a total set $T=\left\{ B_{1},B_{2},...B_{N}\right\} .$ \ A subset of $n\leq N$ of these agents $S_{n}=\left\{ B_{1},...,B_{n}\right\} \subseteq T$ receive $A$'s tweets and also retweet them to $C.$ In other words, $C$ has total of $N$ channels to receive message, but only $n$ of them transfer tweets from $A$. $N$ is a fixed number but $n$ is a variable.

Define $q_{j}$ as the probability that agent $C$ pays attention to
the tweets from agent $B_{j}\in T,$ and we require $\sum_{j=1}^{N}q_{j}=1$.
The joint probability that agent $B_{j}$ is
selected for attention and then agent $C$\ decides to retweets
from $A$ passed along by $B_{j}$ equals to $q_{j}\cdot {\bf P}(C|A;B_{j}).$ Finally,
for some set of agents $S_{n},$ the probability that agent $C$
retweets from agent $A$ through any one of the members in $S_{n}$
is obtained from the \textit{sum of path probabilities} and also known as
the \textit{law of total probability} in CPT:

\begin{equation}
{\bf P}\left( C|A;S_{n}\right) =\sum\limits_{B_{j}\in S_{n}}q_{j}\cdot {\bf P}(C|A;B_{j}).
\label{eq:classic_nq}
\end{equation}

For simplicity we use ${\bf P}(n)$ to represent ${\bf P}( C|A;S_{n})$. In order to calculate Equation~\ref{eq:classic_nq}, we need to know $q_{j}$ and ${\bf P}(C|A;B_{j})
, j=1,2,...n$. 
Actually $q_{i}$ are free parameters which can be hardly estimated from our data. 
$N$ is the maximum number of channels C can pay attention to, which satisfy that ${\bf P}(N+k) = {\bf P}(N), k\in \mathbb{N}^{+}$. 
We consider the initial values ${\bf P}(1)$ can be directly observed from the real data because the interference occurs only when $n>1$. 
In addition, we assume that  $q_{i}=\frac{1-q_{1}}{N-1},i=2,3,...,n$, and ${\bf P}(C|A;B_{j}) = {\bf p}$ for all channels. Finally, the \emph{q-attention} model can be represented by a piecewise defined function:
\begin{equation}
\left\{\begin{matrix}
\mathbf{P}(n) = \mathbf{P}(1) + \frac{1-q_{1}}{N-1} {\bf p}\cdot n, n=2,3,...,N-1\\ 
\mathbf{P}(N+k) = \mathbf{P}(N), k\in \mathbb{N}^{+}
\end{matrix}\right.
\label{eq:linear}
\end{equation}
Now our task is to estimate the parameters of \emph{q-attention} model in order to calculate ${\bf P}(n)$ in Equation~\ref{eq:linear}. Before reaching to the plateau, ${\bf P}(n)$ is just a monotone increasing linear function, with positive slope $\frac{1-q_{1}}{N-1} {\bf p}$ and through a fixed initial point ${\bf P}(1)$. Actually, it is very difficult to estimate $q_{1}, N$ and ${\bf p}$ from real data. Thus, instead of estimating those parameters separately, we will use linear regression approach based on least square principle through a fixed point ${\bf P}(1)$ to fit a linear curve based on data points from the real Twitter data. Then the curve can be used to predict ${\bf P}(n)$.

It is important to note that according to this classic model, ${\bf P}(C|A;S_n)$ increases monotonously with the variable $n$. In other words, adding more channels can never decrease the probability.
\section*{Results}
We based our calculation on two tweets datasets using different crawling methods. One dataset is used to check the information transfer from a global scale, since the tweets come from public timelines. By contrast, the other dataset serves as a case study of information transfer from a local scale, since the tweets comes from an ego network. 

\subsection*{Dataset1}
\begin{figure}[!ht]
\begin{center}
\includegraphics[width=4in]{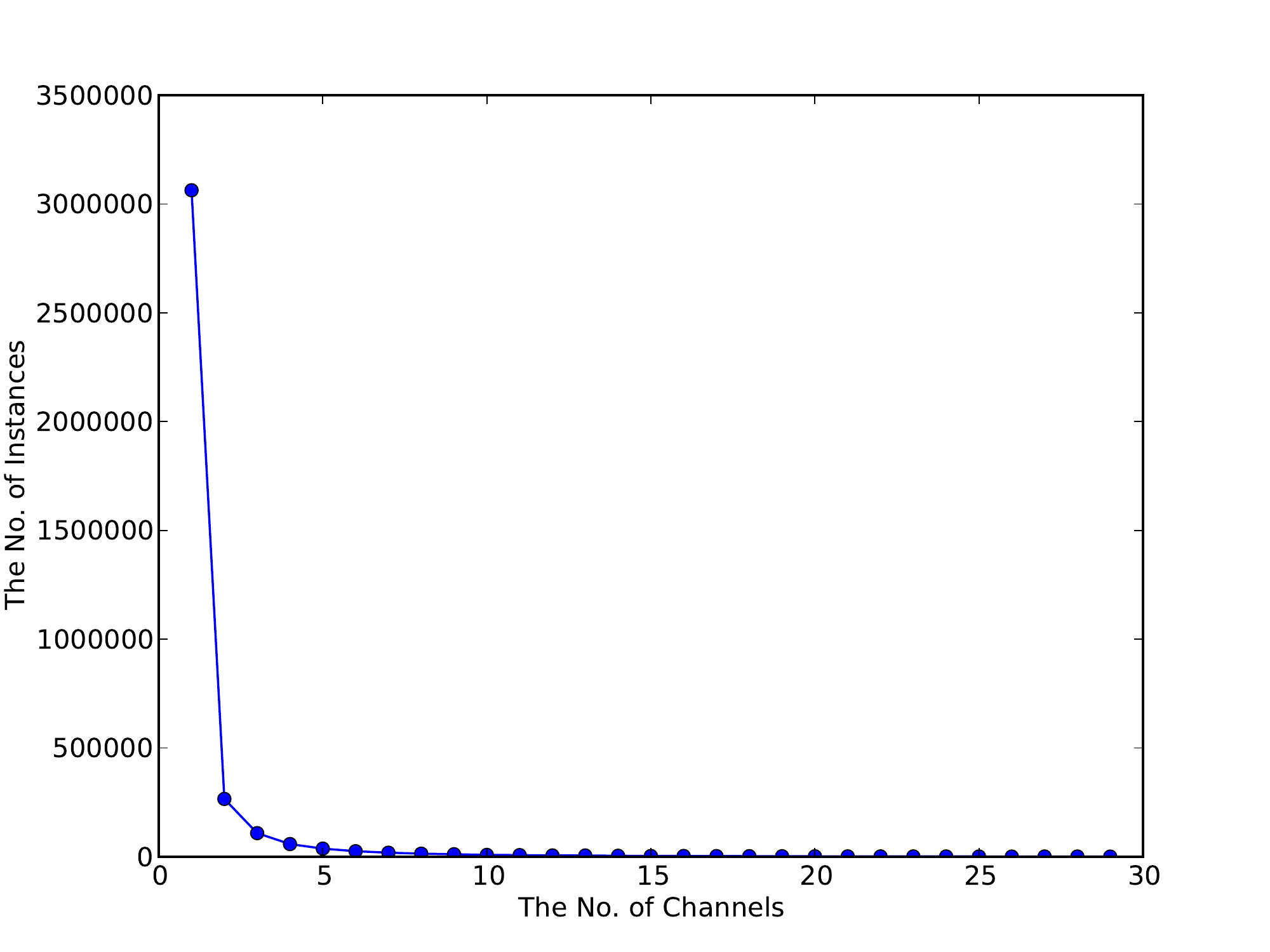}
\end{center}
\caption{
\bf {No. of instances in different information transfer patterns in dataset1}
}
\label{fig:jure1}
\end{figure}

The first dataset contains 467 million tweets from 20 million public users for the time period from June to December 2009 \footnote{http://snap.stanford.edu/data/\#twitter}, which covers 20\%-30\% of total public tweets published on Twitter during that time frame. Only those tweets containing \emph{RT @username1: RT @username2} are selected. Then we grouped those tweets into different patterns based on the number of channels (See Figure~\ref{fig:pattern}). Figure~\ref{fig:jure1} shows the distribution of the amount of information transfer patterns with different number of channels in dataset1. The distribution is skewed towards the information transfer patterns with relatively small number of channels. As the number of channels increases, the number of instances drops significantly. It implies that the occurrence of patterns with a large number of channels is relatively rare.

The largest number of channels we found in our data is 29. However,  we only selected one-channel through six-channel patterns to present our results, for two reasons:
\begin{itemize}
\item The occurrence of patterns with a large number of channels is relatively rare. When the number of channels become very large, their results can be subject to random disturbance which yields unreliable observations.
\item The process of information transfer, described in \emph{q-attention} model, consists of the linear increasing phase and the plateau phase. Only the increasing phase is relevant to the problem under study.

\end{itemize}
The retweeting probability of an instance of n-channel pattern in Twitter data is calculated by:
\begin{equation}
{\bf P}(n) = \frac{\sum \limits_{B_{j}\in S_{n}} No.~of~ tweets~containing~`RT@B_j: RT @A'}{No.~of~tweets~A~posted}
\end{equation}

\begin{figure}[!ht]
\begin{center}
\includegraphics[width=4in]{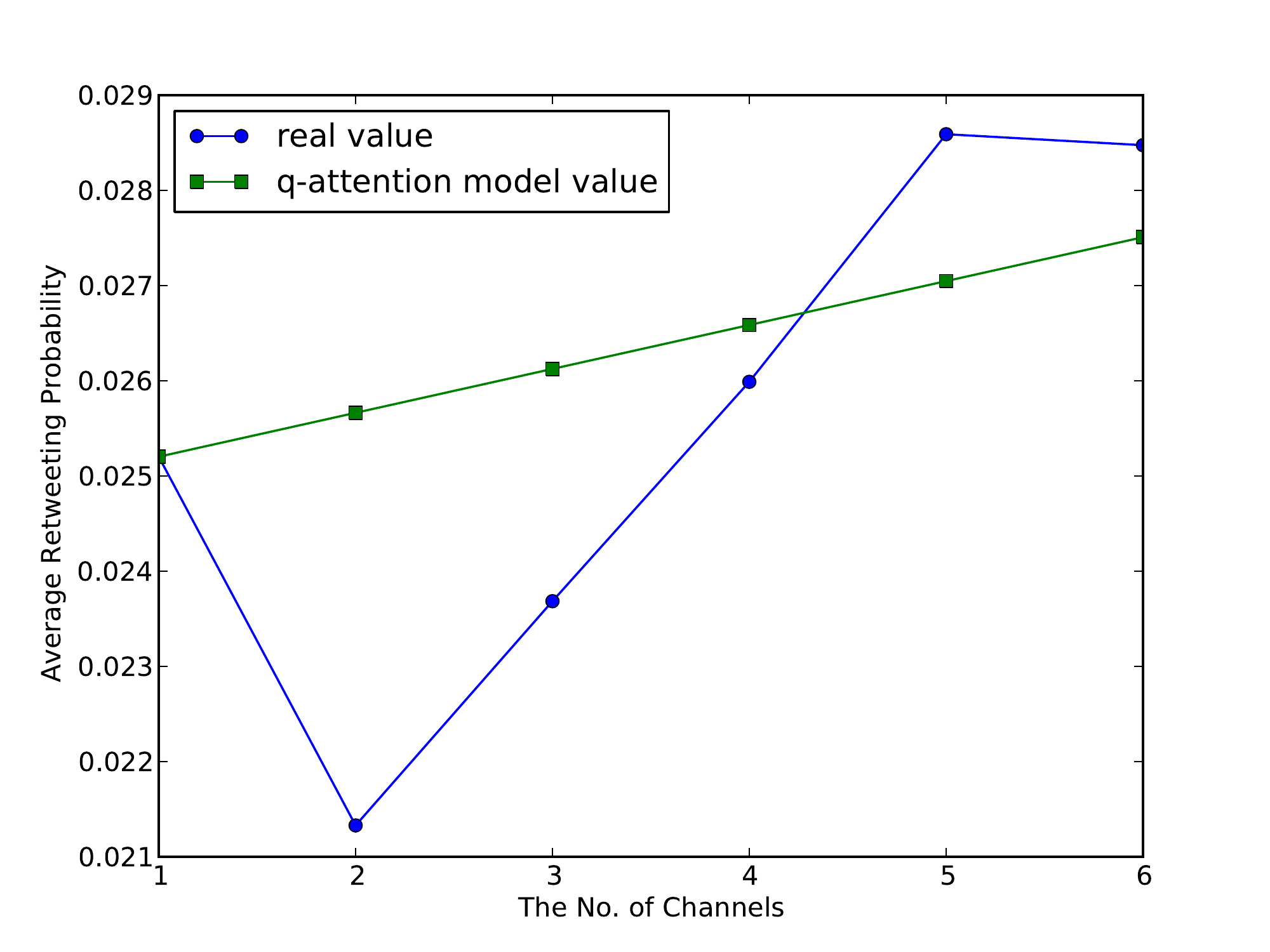}
\end{center}
\caption{\bf {Comparison between real value and classic model in dataset1}}
\label{fig:jure2}
\end{figure}

To present the real retweeting probability for each transfer pattern, we take the average value of retweeting probabilities calculated from all instances belonging to that transfer pattern. And the \emph{q-attention} model results are just the linear regression of the real retweeting probabilities of different patterns. Figure~\ref{fig:jure2} compares the retweeting probability from the real data (noted as real value), with that from the \emph{q-attention} model. Surprisingly, there are two drops in retweeting probability as the number of channels increases: a big one from 1 to 2, and a small one from 5 to 6. It is totally acceptable that the our \emph{q-attention} model value cannot exactly match the real value because of the statistical errors. However, the observation that the amount of information occasionally decrease as the number of channels increase cannot be attributed to pure statistical errors, because it is contradictory to the one of the main conclusions in SIT, that increasing channels always increase the capacity of information being transferred. 
The results from dataset1 provide the global view about information transfer in Twitter, which \emph{q-attention} model based on CPT cannot explain.  

\subsection*{Dataset2}
The second dataset  was crawled by members in Knowledge Engineering Lab at Tsinghua University. This dataset covered time frame from Aug. 1st to Dec. 12th, 2009. Crawling started from a specific user (@yanglicai, a relatively popular user in Chinese Twitter community) of Twitter, and algorithm was designed to check all his contacts involved in replies and retweets, who were not necessarily his followers or followees. Crawling proceeds in a breadth-first approach and results in 192,999 tweets from 8,254 users, and 25.5\% of all tweets collected are retweeting messages. We can view dataset2 as an ego network with @yanglicai in the center and all other nodes having interactions with @yanglicai. 
\begin{figure}[!ht]
\begin{center}
\includegraphics[width=4in]{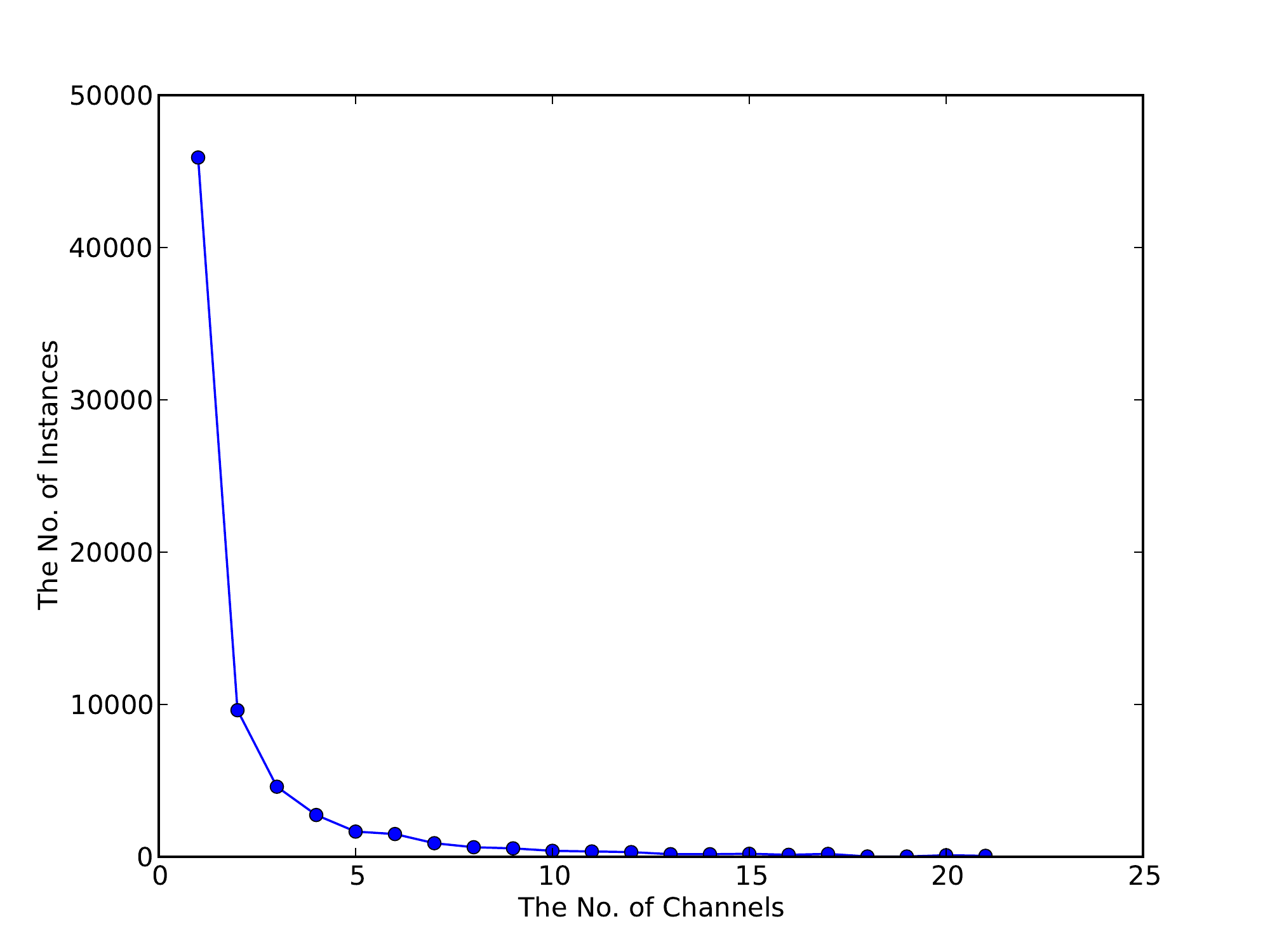}
\end{center}
\caption{
\bf {No. of instances in different information transfer patterns in dataset2}
}
\label{fig:yangzi1}
\end{figure}

\begin{figure}[!ht]
\begin{center}
\includegraphics[width=4in]{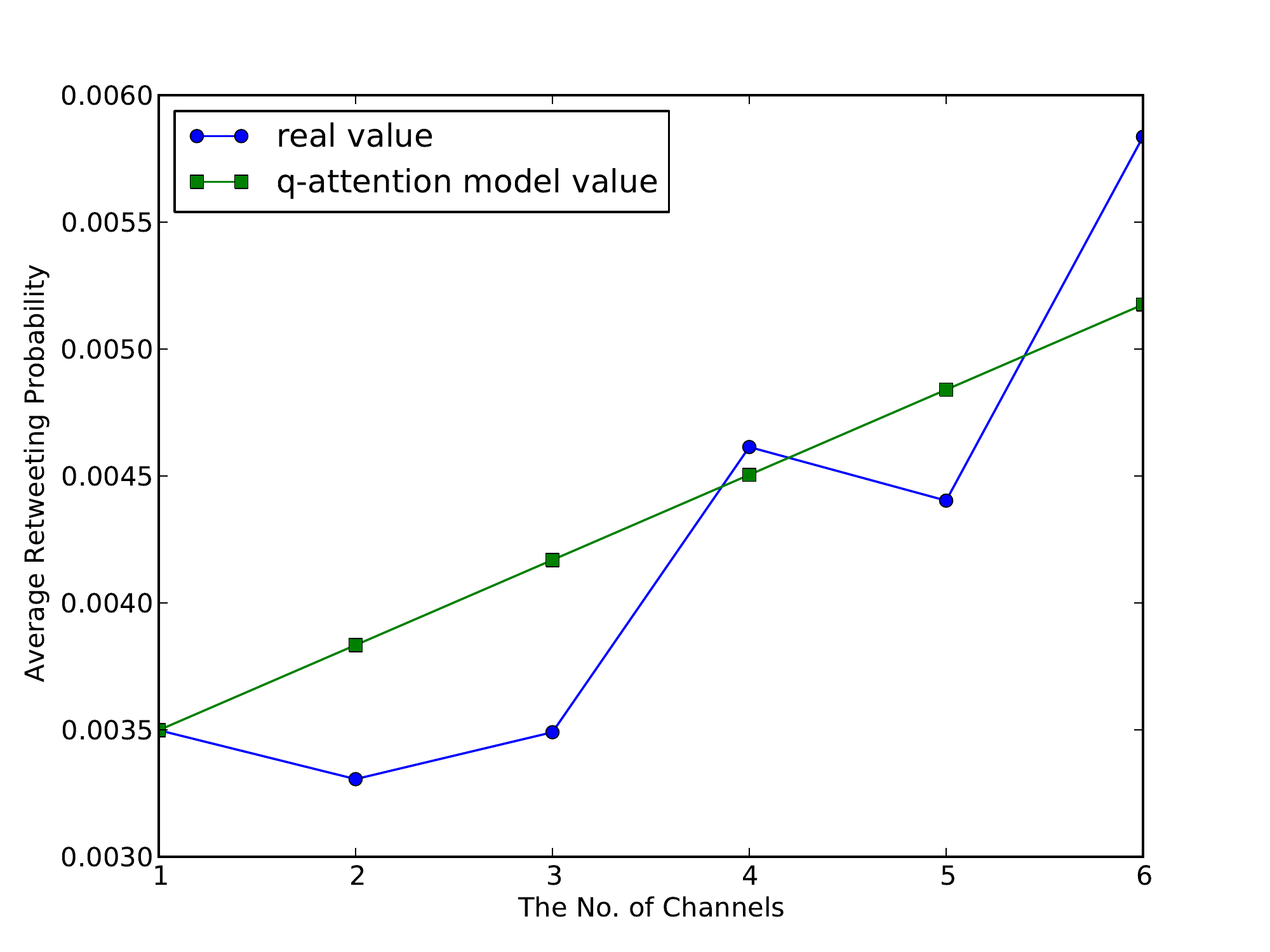}
\end{center}\caption{\bf {Comparison between real value and classic model in dataset2}}
\label{fig:yangzi2}
\end{figure}

Figure~\ref{fig:yangzi1} shows that the average number of records in dataset2 is much smaller than dataset1, because the tweets in dataset2 come from a special ego social network. The largest number of channels we found in dataset2 is 21 and we also selected one-channel to six-channel for analysis. In Figure~\ref{fig:yangzi2}, the difference of retweeting probability between the real data (noted as the real value), and the \emph{q-attention} model is shown. Similar to dataset1, two drops are spotted: from pattern1 to pattern2, from pattern4 to pattern5. The results from dataset2 provide the local scale about information transfer in Twitter, which further confirm the contradictory phenomenon in information transfer as we saw in dataset1 from the global scale.
  
Based on the results from the above two datasets, we found that the amount of information transfer drops occasionally even though the number of information transfer channels increases. It is contradictory to SIT and CPT, which assumes that the increase of channels never decrease the information transfer. Someone may challenge our observation since the \emph{q-attention} model is only one of many approaches to model information transfer and we take several assumptions to further simplify our model into a lineal function. Admittedly, using different classic model may yield different results. However, as long as the model is built on \emph{the law of total probability} from CPT, the result will be consistent with the conclusion that increasing channels never decrease the information transfer amount proved in SIT. Therefore, the contradictory phenomenon we found in Twitter data is independent of which model we use.

\section*{Discussion}
The main deficit of our proposed \emph{q-attention} model lies in the assumption that the $n$ information transfer channels are independent. However, such assumption tends to be counter intuitive. Let's see a concrete example. Assuming that $C$ is a teacher and $A$ is his student. $C$ wants to write a personal evaluation about $A$ and hope to know after-school behaviors about $A$. First, $C$ obtains some of $A$'s information via one of $A$'s classmates $B_1$. Then, in order to know more about $A$, $C$ decides to inquiry $A$'s another classmate $B_2$.

Assuming that the amount of information $C$ gets from $A$ through each single channel, i.e. from B1 or B2, is the same, can we definitely claim that $C$ obtains more information about $A$ after inquiring $B_2$? Not necessary. For example, if $C$ asks $B_1$ whether $A$ has a sister and $B_1$ says \emph{yes}, then $C$ tends to judge $A$ has a sister. However, if $C$ asks $B_2$ the same question but $B_2$ says \emph{no}, then how could $C$ make his judgement? As a result, $C$ may be confused and won't know whether $A$ has a sister or not. In this case, for $C$, the information about $A$ actually decreased when the number of information channels increases from one to two. The reason is that the information provided by two channels is not independent but have mutual interference in semantic. Moreover, such interference not only exist in the semantic aspect but also in the contextual aspect. Specifically, \emph{who} provides the information, as well as \emph{when} and \emph{where} that information is provided, all contribute to the interference effect. QT provides an approach to model such complex inference which beyond the scope of SIT and CPT that only consider the quantity of information but ignore its semantic and contextual aspects. 

Next we will introduce the quantum interference phenomenon in a physics experiment and how we can make an analogy between quantum interferences in physics and in information transfer. Then we will propose a quantum version of \emph{q-attention} model to mathematically interpret the conflict between the increased number of channels and decreased amount of information transfer.

\subsection*{From Double-slit Experiment to Information Transfer}
\begin{figure}[!ht]
\begin{center}
\includegraphics[width=4in]{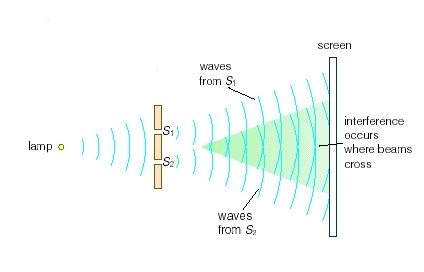}
\end{center}
\caption{
\bf {The double-slit experiment}  
}
\label{fig:double}
\end{figure}

Quantum interference was first observed in a double-slit experiment by Thomas Young in 1803~\cite{young}. Small quantum particles or waves pass through two slits interfered with each other and generated a pattern of bright and dark bands. The beams emerging from the two slits are coherent as they come from the same source (Figure~\ref{fig:double}).

Now we introduce the social information transfer in the context of double slit experiment in Figure~\ref{fig:double}. Image that particles are shot from the lamp towards the two slits $S_1$ and $S_2$. Once the particle passes through $S_1$ or $S_2$, they hit the detector panel, positioned behind the screen, in a particular location $x$ with probability $P_{S_1S_2}(x)$. By closing slit $S_2$, it is possible to measure the probability of particles being detected in position $x$ passing through $S_1$, namely $P_{S_1}(x)$. Similarly, by closing just slit $S_1$, we can measure $P_{S_2}(x)$. Based on CPT, we would expect that the probability particles being detected at $x$ when both slits are open is the sum of the probability of passing through A and being detected at x, $P_{S_1}(x)$, and the probability of passing through B and hitting the detector panel in $x$, $P_{S_2}(x)$. Let's make an analogy between particle shooting and information transfer defined in SIT. The lamp is equivalent to information source, the particles are equivalent information, the two slits are equivalent to two information channels, and the position $x$ in the screen is equivalent to information receiver. The amount of information transferred from the source to $x$ is measured by probability. According to SIT, the total capacity of information transfer is the sum of capacity of each channel, i.e. $P_{S_1S_2}(x)=P_{S_1}(x)+P_{S_2}(x)$. Although the information is actually measured by \emph{bit} in SIT, the \emph{law of total probability} still holds. 

If a third slit $S_3$ is open and particles can also reach $x$ in the panel through it. To measure $P_{S_3}(x)$, we need to close both $S_1$ and $S_2$. As a result, the total information transferred will be $P_{S_1S_2S_3}(x)=P_{S_1}(x)+P_{S_2}(x)+P_{S_3}(x)$. We defined a function $P_n(x)$ representing the information transfer probability with the number of channels. Obviously, $P_n(x)$ is a monotone increase function. However, there should exist a upper limit upon the total amount of information transfer, because not all slits on the wall can let particles pass through and reach position $x$. Similarly, any real information receiver is subject to some receiving limit that increasing the information channel cannot further increasing the information amount as long as the limit is reached. In the \emph{q-attention} model, such limit is considered as human's capacity of paying attention to a finite number of channels. 

As we mentioned above, the monotonous increase property of $P_n(x)$ is built on CPT and confirmed in SIT. However, the double-slit experiment demonstrates the existence of quantum interference effect. In other words, $P_{S_1S_2}(x)\neq P_{S_1}(x)+P_{S_2}(x)$. In particular, $P_{S_1S_2}(x)$ can be zero even though $P_{S_1}(x)>0$ and $P_{S_2}(x)>0$.

%
Now let's see how quantum-like interference may possibly occur in social information transfer. In Figure~\ref{fig:a}, the information is transferred from source $A$ to receiver $C$ via one channel connecting through $B$. There can be no quantum interference effect since only one information channel exists. $C$ has no other alternative channel to obtain $A$'s information except for $B$. However, when there is more than one channel, interference among these channels becomes possible. For example, in Figure~\ref{fig:b}, $C$ can obtain information about $A$ from either $B_1$ or $B_2$. According to SIT, path $A$--$B_1$--$C$ and $A$--$B_2$--$C$ are independent. In other words, the amount of information $C$ obtained from $A$, is just the sum of amount of information form $A$ to $B_1$ to $C$ as if $B_2$ does not exist, and that of from $A$ to $B_2$ to $C$ as if $B_1$ does not exist. Because it assumes that $B_1$ and $B_2$ are totally unrelated from $C$'s perspective. This is also consistent with SIT: more channels always tend to provide more information and therefore reduce uncertainty (entropy) about an event. However, we believe it is oversimplified in social network because the semantic of information are neglected. Assuming that the information content from $B_1$ and $B_2$ are contradictory, then $C$'s uncertainty about $A$ will only increase instead of decreasing. In other words, the total information transferred from $A$ to $C$ actually decreases according to $C$'s cognition, when more channels are available. Here the Ôquantum interferenceÕ takes effect, because $C$ tends to compare the information from different channel and make decisions. In addition, by considering the context of information channels, $C$ is more likely to trust the channel which exerts more social influence upon him. In essence, the SIT only focuses on the quantity of information being transferred but ignore its semantic and contextual perspectives.

\subsection*{Quantum Q-attention Model}

Since the \emph{q-attention} model based on CPT cannot interpret the conflicts between \emph{increasing channel} and \emph{decreasing information transfer}, now we formulate a quantum version of the \emph{q-attention} model, by taking into account the possible interference between channels. According to the probability defined in QT~\cite{stanley}, a probability $p$\
of an outcome is not primitive but derived from
something more primitive called a probability amplitude $\psi $ (a complex
number) -- the probability is obtained by squaring the magnitude of the
amplitude, $p=\left\vert \psi \right\vert ^{2}$. There are three differences between the quantum \emph{q-attention} model and the classic
version.

The first difference is that the classic probability of paying attention to agent $%
B_{j}$, previously denoted $q_{j},$ is replaced by a probability amplitude
denoted $\psi _{j}$. The agent $C$\ is assumed to be in a superposition
state, denoted by 
\begin{equation}
|S_{n}\rangle =\sum_{j=1}^{N}\psi _{j}\cdot |B_{j}\rangle ,
\end{equation}%
which represents this agent's potential to consider any one of the tweets
from agents $B_{j}$ in $T,$ especially including the tweets posted from agent 
$A$ passed along by the agents $B_{j}$ in $T.$ The
probability amplitude $\psi _{j}$ represents the `potential' to consider the
tweets of agent $B_{j},$ but this potential is represented by a complex
number; the probability of paying attention to the tweets of agent $B_{j}$ is
obtained from the squared magnitude $q_{j}=\left\vert \psi _{j}\right\vert
^{2}$ and again we require $\sum_{j=1}^{N}\left\vert \psi _{j}\right\vert
^{2}=1.$

The second difference is that the classic probability that $C$ retweets the
from $A$ through $B_{j}$, previously denoted ${\bf P}(C|A;B_{j})$ is
replaced by the probability amplitude $\langle C|A;B_{j}\rangle .$ The
probability amplitude $\langle C|A;B_{j}\rangle $ represents the `potential'
to retweet, but this potential is represented by a complex number; the
probability of retweeting is obtained from the squared magnitude $%
{\bf P}(C|A;B_{j})=\left\vert \langle C|A;B_{j}\rangle \right\vert ^{2}$ and once
again we require ${\bf P}(C|A;B_{j})+{\bf P}(\bar{C}|A;B_{j})=1.$


The third difference is that QT obeys the \emph{law of total amplitude}
rather than the \emph{law of total probability} from CPT. \ We define $\langle
C|S_{n}\rangle $ as the probability amplitude that the agent $C$\ retweets
the message from $A$ by considering tweets from agents $B_{j}$ in $T.$ The
probability that agent $C$ retweets from agent $A$ passed along
by agents $B_{j}$ in $S_{n}$ then equals ${\bf P}(C|A;S_{n})=\left\vert \langle
C|A;S_{n}\rangle \right\vert ^{2}.$ According to QT, to determine
the amplitude $\langle C|A;S_{n}\rangle $, we replace the sum of path
probabilities shown in Equation \ref{eq:classic_nq} with the sum of path
amplitudes given below%
\begin{equation}
\langle C|A;S_{n}\rangle =\sum_{B_{j}\in S_{n}}\psi _{j}\cdot \langle
C|A;B_{j}\rangle .  \label{Law of total amplitude}
\end{equation}


\subsection*{Predictions for n=1,2,3}

Let us see some examples of how our \emph{q-attention} model encompass the interference effect. First consider $n=1$ in \ which case there is
only one agent $S_{1}=\left\{ B_{1}\right\} .$ \ Then 
in this case $%
{\bf P}(1)=|\psi _{1}\cdot \langle C|A;B_{1}\rangle |^{2}=\left\vert \psi
_{1}\right\vert ^{2}\cdot \left\vert \langle C|A;B_{1}\rangle \right\vert
^{2}. $ The probability that $C$ retrweets $A$ through this one agent 
is virtually the same as the classic version and
no quantum interference occurs in this case.

Next consider $n=2$ in which case there are two agents $S_{2}=$ $\left\{
B_{1},B_{2}\right\} .$ 
In this case Equation \ref{Law of total
amplitude} produces 
\begin{eqnarray*}
{\bf P}(2) &=&\left\vert \psi _{1}\cdot \langle C|A;B_{1}\rangle +\psi
_{2}\cdot \langle C|A;B_{2}\rangle \right\vert ^{2} \\
&=&\left\vert \psi _{1}\cdot \langle C|A;B_{1}\rangle \right\vert
^{2}+\left\vert \psi _{2}\cdot \langle C|A;B_{2}\rangle \right\vert ^{2} \\
&&+\left( \psi _{1}^{\ast }\cdot \psi _{2}\right) \cdot \langle
C|A;B_{1}\rangle ^{\ast }\cdot \langle C|A;B_{2}\rangle \\
&&+\left( \psi _{1}\cdot \psi _{2}^{\ast }\right) \cdot \langle
C|A;B_{1}\rangle \cdot \langle C|A;B_{2}\rangle ^{\ast }.
\end{eqnarray*}%
The first two terms correspond to the same probabilities that one would
obtain from the classic model. The last two terms form a conjugate pair, 
\begin{eqnarray*}
\left( \psi _{1}^{\ast }\cdot \psi _{2}\right) \cdot \langle C|A;B_{1}\rangle
^{\ast }\cdot \langle C|A;B_{2}\rangle &=&\left\vert \psi _{1}\cdot \psi
_{2}\cdot \langle C|A;B_{1}\rangle \cdot \langle C|A;B_{2}\rangle \right\vert
\cdot \left( \cos \left( \theta _{12}\right) +i\cdot \sin \left( \theta
_{12}\right) \right) \\
\left( \psi _{1}\cdot \psi _{2}^{\ast }\right) \cdot \langle C|A;B_{1}\rangle
\cdot \langle C|A;B_{2}\rangle ^{\ast } &=&\left\vert \psi _{1}\cdot \psi
_{2}\cdot \langle C|A;B_{1}\rangle \cdot \langle C|A;B_{2}\rangle \right\vert
\cdot \left( \cos \left( \theta _{12}\right) -i\cdot \sin \left( \theta
_{12}\right) \right) ,
\end{eqnarray*}%
where $\theta _{12}$ is the phase of each complex number which depends on
the pair of agents $\left\{ B_{1},B_{2}\right\} .$ The sum of this conjugate
pair produces a real number called the interference term:%
\begin{equation*}
Int_{12}=2\cdot \left\vert \psi _{1}\cdot \psi _{2}\cdot \langle
C|A;B_{1}\rangle \cdot \langle C|A;B_{2}\rangle \right\vert \cdot \cos \left(
\theta _{12}\right) .
\end{equation*}%
The cosine term can be positive (producing constructive interference),
negative (producing destructive interference), or zero (producing no
interference). If the cosine is sufficiently negative, then the probability
of $C$ indirectly retweets from A when given two paths can be smaller than the
probability given a single path. \ This happens whenever $\left\vert
\psi _{2}\cdot \langle C|A;B_{2}\rangle \right\vert ^{2}+Int<0.$ 


When there is no interference, $\cos \left( \theta _{12}\right) =0,$ then
the probability that $C$ retweets from A based on two intermediate
agents must be larger than the probability based on a single agent. This
follows the fact that when there is no interference, the probabilities
from each path sum and exceed the single path just like the classic
probability model. \ 

Now consider the case with $n=3$ and $S_{3}=\left\{
B_{1},B_{2},B_{3}\right\} .$ \ 
In this case we obtain%
\begin{eqnarray*}
{\bf P}(3) &=&\left\vert \psi _{1}\cdot \langle C|A;B_{1}\rangle +\psi
_{2}\cdot \langle C|A;B_{2}\rangle +\psi _{3}\cdot \langle C|A;B_{3}\rangle
\right\vert ^{2} \\
&=&\left\vert \psi _{1}\cdot \langle C|A;B_{1}\rangle \right\vert
^{2}+\left\vert \psi _{2}\cdot \langle C|A;B_{2}\rangle \right\vert
^{2}+\left\vert \psi _{3}\cdot \langle C|A;B_{3}\rangle \right\vert ^{2} \\
&&+Int_{12}+Int_{13}+Int_{23},
\end{eqnarray*}%
with the interference defined as 
\begin{equation*}
Int_{ij}=2\cdot \left\vert \psi _{i}\cdot \psi _{j}\cdot \langle
C|A;B_{i}\rangle \cdot \langle C|A;B_{j}\rangle \right\vert \cdot \cos \left(
\theta _{ij}\right) .
\end{equation*}%
Each pair of cosines can be positive or negative. If they are all negative,
then we could find a decrease in probability for three paths as compared to
two paths or one path. But if the cosines flip from positive to negative,
they could cancel out in the sum leaving the result with no overall
interference. In general, as more path amplitudes are added together there
is a tendency for the interference terms to change sign and cancel out,
which is called decoherence. When decoherence occurs, the quantum model
starts behaving like the classic model.


\section*{Related Work}
QT has been applied in information retrieval~\cite{van}. Furthermore, research demonstrates that quantum-like phenomena exist in human natural language, cognition and decision making~\cite{peter, peter2}. For example, the term ÒbatÓ can be modeled in a two dimensional vector space as a vector representing a linear combination of two senses: ÒsportsÓ and ÒanimalÓ. The same goes for the term ÒboxerÓ: ÒsportsÓ and ÒanimalÓ. Then the combination of Òboxer batÓ follows the quantum formalism~\cite{peter}. Experiments show that many conceCPT combinations are non-separable meaning that the combination cannot be modeled by probability distributions across the senses of the individual words. The non-linear or non-separable combination of two entities exists not only in information retrieval but also other fields which deal with social interaction and cognitive interference. Phenomena containing interference cannot be modeled via classical probabilistic models. QT offers a probabilistic, logic and geometric formalism based on the mathematics of Hilbert spaces to describe the behavior of interference. 

Van Rijsbergen~\cite{van2} provided a mechanism to describe logic-based IR models within a quantum formalism, in particular with Hilbert spaces.  One of the well-adopted is Logical Imaging (LI)~\cite{zuccon}. LI calculates a probability that a document is relevant to a query by considering the correlation of terms that appears across different documents. Huertas-Rosero, Azzopardi, and van Rijsbergen~\cite{zuccon2} proposed quantum-based measures for documents by using the selective erasers. Erasers model the relation of one term with respect to another by using co-occurrence methods, and the distance between this term and other neighboring occurrences. Arafat and van Rijsbergen~\cite{van} applied QT to address some fundamental issues in search, which investigate and model user cognition and their interactions during the search process. Di Buccio, Lalmas, and Melucci~\cite{buccio} proposed a uniform way to model properties of entities, relationships of entities, and properties of relationships through a geometric framework in terms of vector subspaces~\cite{massimo}. Zuccon, Azzopardi and van Rijsbergen~\cite{zuccon2} proposed to represent documents and queries as subspaces rather than as vectors in traditional vector-space model. Then the relevance between the document and user query can be reformulated as how to distinguish preparations of different quantum systems, which can be measured using subspace distance. Hou and Song~\cite{yuexian} proposed an extended vector space model (EVSM) to model context-sensitive high-order information. Zuccon, Azzopardi, and van Rijsbergen~\cite{zuccon2} suggest to rank the document relevance using QT as judgment of relevant is not independent from other documents, and the interference of other documents play an important role to judge the relevance.  They proposed a novel quantum probability ranking principle (QPRP) to model situations that a document relevance assessment is influenced by other documents. Interference exists in document relevance judgment, especially when users change their relevance measurement for one document after they have measured other documents. Zhang et al.~\cite{zhang} proposed to use quantum finite automation (QFA) to represent the transition of the measurement states (the relevance degrees of the document judged by users) and dynamically model the cognitive interference of users when they are influenced by other related documents. Piwowarski, Frommholz, Lalmas, and van Rijsbergen~\cite{piwo} applied tensor products from QT to refine the representation of queries (as density operators) and documents (as subspaces). Although QT has been extensively applied to IR and human cognition modeling, but none have apply QT to explain the information transfer in social networks, such as Twitter.

\section*{Conclusion and Future Work}
In this paper, we studied whether quantum-like phenomena can be spotted in social networks. To address it, we constructed two social networks formed by the retweeting relationship in Twitter, proposed a CPT based \emph{q-attention} model to quantify the information transfer by retweeting probability, and compared the model results with the real value. We found that CPT and SIT cannot interpret the conflict between decreased information transfer and increased channels and proposed a quantum version of \emph{q-attention} model to solve the conflict.

Although QT provides some promising potential to add semantic and context to classic information transfer, a great deal of future work remains. First, we should select more social network datasets, like Facebook, Myspace, or Blogs, to test the quantum-like phenomenon. Second, the quantum model we proposed only gives the theoretical possibility to explain the quantum-like interference, but does not provide empirical interpretation of what the interference are. We may need to apply the social influence theory to further explain it. Third, some psychological experiments can be conducted to better understand the quantum-like phenomenon.

\bibliography{refs}



\end{document}